\documentclass[12pt, a4paper] {article}

\usepackage{amsfonts}
\usepackage{verbatim}
\usepackage{color}
\usepackage{amsmath}
\usepackage{graphicx}
\usepackage{bm}
\usepackage{amssymb}
\usepackage{latexsym}
\usepackage{mathrsfs}
\usepackage{textcomp}
\begin{document}

\begin{center}
{\bf\large{Faddeev-Jackiw Quantization of Christ-Lee Model} }

\vskip 1.5 cm

{\sf{ \bf Anjali S and Saurabh Gupta}}\\
\vskip .1cm
{\it Department of Physics, National Institute of Technology Calicut,\\ Kozhikode - 673 601, Kerala, India}\\
\vskip .15cm
{E-mails: {\tt anjalisujatha28@gmail.com, saurabh@nitc.ac.in}}
\end{center}
\vskip 1cm

\noindent
{\bf Abstract:}
We analyze the constraints of Christ-Lee model by the means of modified Faddeev-Jackiw formalism in Cartesian 
as well as polar coordinates. Further, we accomplish quantization {\it \`{a} la} Faddeev-Jackiw by choosing appropriate 
gauge conditions in both the coordinate systems. Finally, we establish gauge symmetries of Christ-Lee model with 
the help of zero modes of the symplectic matrix.

\vskip 1.5 cm



\newpage

\section{Introduction}
The conventional method of quantization is not directly applicable to the dynamical systems with singular Lagrangian or systems 
embedded with inherent constraints. Dirac proposed a formalism for such systems and introduced the concept 
of Dirac brackets. Dirac formalism also introduces the classification of constraints into different classes such as 
primary, secondary, tertiary, etc. and further into first-class and second-class \cite{1}. Although this formalism does the 
designated task, but it is found to be tedious in some cases. 

However, an alternative approach, which is geometrically motivated and make use of the symplectic structure of 
phase space,  was introduced by Faddeev and Jackiw \cite{2}. In this formalism, all the constraints are treated 
on equal footing without any further classification. One of the main requirements of this formalism is the necessity 
of first-order Lagrangian describing the system. 
The constraints are obtained from zero modes of the symplectic two-form matrix and further they are incorporated 
in the Lagrangian in an iterative manner till all the constraints in the theory are eliminated (see, e.g. \cite{3,4}). The brackets among basic 
fields, obtained from the inverse of the non-singular symplectic two-form matrix, coincides with that procured from the Dirac formalism \cite{25}. 
In the modified Faddeev-Jackiw formalism the consistency condition of constraints along with the symplectic equations of motion is used to derive 
new constraints (cf. \cite{23,24,5,6} for details).

The Faddeev-Jackiw formalism of symplectic analysis is applied in various cases such as four dimensional Pontryagin 
and Euler invariant \cite{7},  four dimensional BF theory \cite{8}, topologically massive AdS
gravity \cite{9}, three dimensional (non-)Abelian exotic action for gravity \cite{11} and also in higher-derivative theories \cite{26}. Furthermore, 
Faddeev-Jackiw quantization for constrained system is also implemented in path integral framework \cite{10}.

On the other hand, Christ-Lee model is one of the simplest examples of a singular system described in the classical regime in three 
dimensional space \cite{13,15}. Being a constrained (singular) system, it has been studied in the framework {\it \`{a} la} Dirac 
which shows the existence of a set of two independent first-class constraints \cite{15}. These first-class constraints, 
in turn, imply that the underlying theory is a gauge theory. As far as quantization of this model is concerned there are two specific 
gauge choices depending upon the coordinate system in which the model is explored \cite{13,15}.

The various aspects of Christ-Lee model has been exclusively explored such as quantization in the framework of path integral 
formulation \cite{14} and using WKB approximation \cite{16}. This model is also being studied in the framework of BRST formalism \cite{15}. 
In one of the recent works, the existence of (anti-)co-BRST symmetries, bosonic and discrete symmetries have been established \cite{17}. 
This, in turn, implies the Christ-Lee model as a simple model for the Hodge theory \cite{17,21,22}. Moreover, these (anti-)BRST  and (anti-)co-BRST 
symmetries are derived within the framework of augmented supervariable approach \cite{18}.

The main motivation behind our present investigation is to explore Christ-Lee model in a geometrically motivated formalism and deduce its 
constraint structure. Second, to quantize this model and procure all the basic brackets within the framework of modified Faddeev-Jackiw formalism. 
Finally, we wish to obtain the gauge transformations and provide physical interpretation to the Lagrange multipliers present in the theory.

The content of this paper is organized as follows. Section 2 deals with the derivation of constraints of Christ-Lee model, in polar 
coordinates, with the aid of modified Faddeev-Jackiw formalism. We also obtain the full set of basic brackets of the theory and 
establish the gauge transformations.
Our section 3 gives a detail account about the constraints and basic brackets of Christ-Lee model, in Cartesian coordinates, within the 
framework of modified Faddeev-Jackiw approach. We also list out the gauge transformations and provide a new 
interpretation for the Lagrange multipliers present in the model. Finally, we summarize our results and furnish some future directions in Section 4.

\section{Christ-Lee Model in Polar Coordinates: Faddeev-Jackiw Quantization}
We begin with the Lagrangian of Christ-Lee model, in polar coordinates, as described by \cite{13}
\begin{eqnarray}
L = \frac{1}{2}\dot{r}^2+\frac{1}{2}r^{2}(\dot{\theta}-z)^{2}-V(r),
\end{eqnarray}
where $\dot{r}$ and $\dot{\theta}$ represents the generalized velocities and $z$ is a generalized coordinate. The canonical conjugate momenta 
corresponding to the generalized coordinates $r,\theta$ and $z$ are given, respectively, as
\begin{eqnarray}
P_{r}= \frac{\partial L}{\partial \dot{r}} = \dot{r},\qquad
P_{\theta} =  \frac{\partial L}{\partial \dot{\theta}} = r^{2}\dot{\theta}-r^{2}z,\qquad
P_{z} = 0.
\end{eqnarray}
The Hamiltonian of the system derived from the above Lagrangian is given by
\begin{eqnarray}
H = \frac{P_{r}^{2}}{2}+\frac{P_{\theta}^{2}}{2r^{2}}+P_{\theta}z+V(r).
\end{eqnarray}
To make use of Faddeev-Jackiw formalism we need to express the above Lagrangian in first-order form. 
Thus, the first-order Lagrangian is \cite{17}
\begin{eqnarray}
L_{f}^{(0)} &=& \dot{r}P_{r}+\dot{\theta}P_{\theta}-V^{(0)}, \label{L_0}
\end{eqnarray}
where $V^{(0)}$ denotes following symplectic potential
\begin{eqnarray}
V^{(0)} &=& \frac{P_{r}^{2}}{2}+\frac{P_{\theta}^{2}}{2r^{2}}+zP_{\theta}+V(r).
\end{eqnarray}
In this formalism, the equations of motion are derived in terms of symplectic matrix $f_{ij}^{(0)}$ in the following manner 
\begin{eqnarray}
f_{ij}^{(0)}\dot{\zeta^{j}} &=& \frac{\partial V^{(0)}(\zeta)}{\partial \zeta^{i}}, \label{fj_eqm}
\end{eqnarray}
where the symplectic matrix is given by
\begin{eqnarray}
f_{ij}^{(0)} &=& \frac{\partial a_{j}(\zeta)}{\partial \zeta^{i}}- \frac{\partial a_{i}(\zeta)}{\partial \zeta^{j}}, \quad 
{\text with} \quad a_{i} = \frac{\partial L}{\partial \dot{\zeta_{i}}}. \label{f_ij}
\end{eqnarray}
Now, the set of symplectic variables are 
\begin{eqnarray}
\zeta^{(0)}= \left\{r,P_{r},\theta,P_{\theta},z\right\}.
\end{eqnarray}
Therefore, the components of symplectic one-form can be computed with the help of (\ref{f_ij}) and are listed below:
\begin{eqnarray}
a^{(0)}_{r}=P_{r}, \quad a^{(0)}_{P_{r}}=0, \quad a^{(0)}_{\theta}=P_{\theta}, \quad a^{(0)}_{P_{\theta}}=0, \quad a^{(0)}_{z}=0.
\end{eqnarray}
Thus, the deduced symplectic matrix $(f_{ij}^{(0)})$ takes following form 
\begin{eqnarray}
f_{ij}^{(0)} =
\begin{pmatrix}
0 & -1 &0 &0 &0 \\
1 &0 &0 &0 &0 \\
0 &0 &0 &-1 &0\\
0 &0 &1 &0 &0\\
0 &0 &0 &0 &0
\end{pmatrix},
\end{eqnarray}
here $f_{ij}^{(0)}$ is a singular matrix. The presence of singular matrix indicates that the system contains constraints. Thus, 
zero-mode of the matrix is calculated as $(\nu^{(0)})^{T}=(0,0,0,0,v^{z})$, where $v^{z}$ is an arbitrary constant. In the view of 
Faddeev-Jackiw formalism, this zero mode will give rise to the constraint in the system:
\begin{eqnarray}
\Omega^{(0)} = (\nu^{(0)})^{T}\frac{\partial V^{(0)}(\zeta)}{\partial \zeta^{(0)}} = 0 \quad \Longrightarrow \quad 
\Omega^{(0)} = \nu^{z}P_{\theta} = 0. \label{11}
\end{eqnarray} 
Now we employ the modified Faddeev-Jackiw method to deduce new constraints in the theory. Here we make use of the consistency condition 
of constraints which is analogous to the `Dirac-Bergmann' approach to derive the same. So, according to the modified formalism, we have
\begin{eqnarray}
\dot{\Omega}^{(0)} = \frac{\partial{\Omega^{(0)}}}{\partial \zeta^{i}}\dot{\zeta^{i}} =0 .\label{mfj_c}
\end{eqnarray}
Combining \eqref{fj_eqm} and \eqref{mfj_c} we obtain
\begin{eqnarray}
f_{kj}^{(1)} \dot{\zeta^{j}} = Z_{k}(\zeta), \label{mfj_eqm}
\end{eqnarray}
where
\begin{eqnarray}
f_{kj}^{(1)}=
\begin{pmatrix}
f_{ij}^{(0)}\\
\frac{\partial{\Omega^{(0)}}}{\partial \zeta^{i}}
\end{pmatrix}
, \qquad Z_{k}(\zeta)=
\begin{pmatrix}
\frac{\partial V^{(0)}(\zeta)}{\partial \zeta^{i}}\\
0
\end{pmatrix}.
\end{eqnarray}
Now calculating the symplectic matrix $f_{kj}^{(1)}$ yields
\begin{eqnarray}
 f_{kj}^{(1)} =
\begin{pmatrix}
0 &-1 &0 &0 &0 \\
1 &0 &0 &0 &0 \\
0 &0 &0 &-1 &0\\
0 &0 &1 &0 &0\\
0 &0 &0 &0 &0\\
0 &0 &0 &1 &0
\end{pmatrix}.
\end{eqnarray}
Even though the matrix $f_{kj}^{(1)}$ is a non-square matrix, it has a zero-mode. The zero-mode of the above matrix is 
$(\nu^{(1)})^{T}=(0,0,1,0,v_{1}^{z},1)$, where $v_{1}^{z}$ is an arbitrary constant. Multiplying $(\nu^{(1)})^{T}$ to \eqref{mfj_eqm}, 
gives the constraints in the model when evaluated with the condition $\Omega^{(0)}=0$. So, we have
\begin{eqnarray}
 (\nu^{(1)})^{T} Z_{k} (\zeta)|_{\Omega^{(0)}=0} = 0 \; \Longrightarrow \; 
\nu_{1}^{z}P_{\theta}|_{\Omega^{(0)}=0} = 0,
\end{eqnarray}
which gives identity. Therefore, there are no further constraints present in the theory. 
Now we introduce the obtained constraint into the first-order Lagrangian using Lagrange multiplier $(\lambda)$, as follows
\begin{eqnarray}
L_f^{(1)}=\dot{r}P_{r}+\dot{\theta}P_{\theta}+\dot{\lambda}P_{\theta}-V^{(1)},
\end{eqnarray}
where
\begin{eqnarray}
V^{(1)} = V^{(0)}|_{P_{\theta}=0} \; = \; \frac{P_{r}}{2}^{2}+V(r).
\end{eqnarray}
The set of first-iterated symplectic variables $(\zeta^{(1)})$ are
\begin{eqnarray}
\zeta^{(1)} = \{r, P_{r}, \theta, P_{\theta}, \lambda\},
\end{eqnarray}
and the respective symplectic one-forms are listed below:
\begin{eqnarray}
a^{(1)}_{r}=P_{r}, \quad a^{(1)}_{P_{r}}=0, \quad a^{(1)}_{\theta}=P_{\theta}, \quad a^{(1)}_{P_{\theta}}=0, \quad a^{(1)}_{\lambda}=P_{\theta}.
\end{eqnarray}
The first-iterated symplectic matrix is calculated accordingly and found to be 
\begin{eqnarray}
f_{ij}^{(1)} =
\begin{pmatrix}
0 & -1 &0 &0 &0 \\
1 &0 &0 &0 &0 \\
0 &0 &0 &-1 &0\\
0 &0 &1 &0 &1\\
0 &0 &0 &-1 &0
\end{pmatrix}, \label{sym1}
\end{eqnarray}
where $f_{ij}^{(1)}$ is still a singular matrix. As there are no further constraints in theory and the symplectic matrix is still singular
indicate that the system has a gauge symmetry.

Thus, in order to quantize the underlying theory we choose the gauge condition $\theta=0$ \cite{13, 15} and introduce it into the Lagrangian
with the help of a Lagrange multiplier $(\rho)$ 
\begin{eqnarray}
L_f^{(2)}=\dot{r}P_{r}+\dot{\theta}P_{\theta}+\dot{\lambda}P_{\theta}+\dot{\rho}\theta-V^{(2)}, \label{pol_L2}
\end{eqnarray}
where
\begin{eqnarray}
V^{(2)}=V^{(1)}|_{\theta=0} \; = \; \frac{P_{r}}{2}^{2}+V(r).
\end{eqnarray}
Now the set of second-iterated symplectic variables are
\begin{eqnarray}
\zeta^{(2)}= \{r, P_{r}, \theta, P_{\theta}, \lambda, \rho\},
\end{eqnarray}
with the symplectic one-forms are given by
\begin{eqnarray}
a^{(2)}_{r}=P_{r}, \quad a^{(2)}_{P_{r}}=0, \quad a^{(2)}_{\theta}=P_{\theta}, \quad a^{(2)}_{P_{\theta}}=0, \quad a^{(2)}_{\lambda}=P_{\theta},
\quad a^{(2)}_{\rho}=\theta.
\end{eqnarray}
Thus, the second-iterated symplectic matrix can be constructed in the following fashion
\begin{eqnarray}
f_{ij}^{(2)} =
\begin{pmatrix}
0 & -1 &0 &0 &0 &0 \\
1 &0 &0 &0 &0 &0 \\
0 &0 &0 &-1 &0 &1\\
0 &0 &1 &0 &1 &0\\
0 &0 &0 &-1 &0 &0\\
0 &0 &-1 &0 &0 &0
\end{pmatrix},
\end{eqnarray}
which is obviously a non-singular matrix. So, its inverse can be calculated as
\begin{eqnarray}
(f_{ij}^{(2)})^{-1} =
\begin{pmatrix}
0 & 1 &0 &0 &0 &0 \\
-1 &0 &0 &0 &0 &0 \\
0 &0 &0 &0 &0 &-1\\
0 &0 &0 &0 &-1 &0\\
0 &0 &0 &1 &0 &1\\
0 &0 &1 &0 &-1 &0
\end{pmatrix}.
\end{eqnarray}
The components of the inverse matrix $(f_{ij}^{(2)})^{-1}$ directly give the basic brackets in the theory. Thus, the basic brackets 
in the theory are
\begin{eqnarray}
&&\{r,P_{r}\} =1=-\{P_{r},r\}, \qquad \{\lambda,P_{\theta}\} =1=-\{P_{\theta},\lambda\},\\ \nonumber
&&\{\rho,\theta\} =1=-\{\theta,\rho\}, \qquad \{\lambda,\rho\} =1=-\{\rho,\lambda\}.
\end{eqnarray}

As we have already mentioned that the first-iterated symplectic matrix $(f_{ij}^{(1)})$, which turns out to be a singular matrix (cf. (\ref{sym1})),  
indicates that the underlying theory is a gauge theory. 
Consequently, it is evident from the symplectic equations of motion that the zero mode of this singular 
matrix is orthogonal to the gradient of potential (cf. \eqref{11} for details). This, in turn, implies that they are generators of local 
displacements on the isopotential surface and hence act as a generators of gauge symmetry \cite{11, mont,nati,omar}. 

Thus, the following zero mode of the symplectic matrix $f_{ij}^{(1)}$ 
\begin{eqnarray}
(\upsilon^{(1)})^{T} = (0,0,1,0,-1),
\end{eqnarray}
acts as a generators of gauge transformations $(\delta)$ in the following fashion
\begin{eqnarray}
\delta\zeta_{k}^{(1)} = \upsilon_{k}^{(1)}\kappa,
\end{eqnarray}
where $\zeta_k^{(1)}$ is the set of all symplectic variables and $\kappa (t)$ denotes the time-dependent infinitesimal gauge parameter. 
Thus, the gauge transformations of the Christ-Lee model, in the polar coordinates, can be deduced as follows 
\begin{eqnarray}
\delta{\theta}=\kappa(t), \qquad \delta{z}=\dot{\kappa}(t), \qquad \delta[{r,P_{r},P_{\theta}}]=0.
\end{eqnarray}
It is straightforward to verify that our first-order Lagrangian $(L^{(0)}_f)$  
remains invariant under this set of gauge transformations.

\subsection{Interpretation of Lagrange Multipliers}
The Lagrange multipliers are introduced as auxiliary fields in Faddeev-Jackiw formalism to incorporate constraints into 
the Lagrangian. A new interpretation for Lagrange multipliers can be provided in this formalism as the basic brackets involving 
these multipliers do not appear in the Dirac formalism \cite{24}. Therefore, by solving symplectic equations of motion, we found 
that these auxiliary coordinates are related with the physical coordinates of the theory.
So, to obtain new information about these Lagrange multipliers, we recall symplectic equations of motion from \eqref{pol_L2}
\begin{eqnarray}
f_{ij}^{(2)}\dot{\zeta}^{(2)j} = \frac{\partial V^{(2)}(\zeta)}{\partial \zeta^{(2)i}},
\end{eqnarray}
\begin{eqnarray} \label{f_ij2}
\begin{pmatrix}
0 & -1 &0 &0 &0 &0 \\
1 &0 &0 &0 &0 &0 \\
0 &0 &0 &-1 &0 &1\\
0 &0 &1 &0 &1 &0\\
0 &0 &0 &-1 &0 &0\\
0 &0 &-1 &0 &0 &0
\end{pmatrix} 
\begin{pmatrix}
\dot{r}\\
\dot{P_{r}}\\
\dot{\theta}\\
\dot{P_{\theta}}\\
\dot{\lambda}\\
\dot{\rho}
\end{pmatrix} =
\begin{pmatrix}
\frac{\partial V}{\partial r}\\
P_{r}\\
0\\
0\\
0\\
0
\end{pmatrix}.
\end{eqnarray}
From above relationship, we have
\begin{eqnarray}
\dot{\lambda}=-\dot{\theta}, \qquad \dot{\rho}=\dot{P_{\theta}}. 
\end{eqnarray}
Therefore, we can infer that $\lambda$ can be represented in  terms of one of the coordinates in the 
theory whereas $\rho$ can be depicted in terms of the momenta corresponding to that coordinate.
Before wrapping up this subsection, we would like to point out that 
these Lagrange multipliers appear as $\dot \lambda$ and $\dot \rho$ in the Lagrangian (cf. \eqref{pol_L2}). 
Therefore, we choose to provide an interpretation in terms of $\dot \lambda$ and $\dot \rho$. 
Alternatively, one can provide an interpretation by expressing them in terms of $\lambda$ and $\rho$ as well, however 
the underlying dynamics remains unaffected.

\section{Christ-Lee Model in Cartesian Coordinates : Faddeev-Jackiw Quantization}
The Lagrangian describing dynamics of the Christ-Lee model in Cartesian coordinate system is given by \cite{13,15}
\begin{eqnarray}
\tilde{L}= \frac{\dot{x}}{2}^{2} + \frac{\dot{y}}{2}^{2} - z(x\dot{y}-y\dot{x}) + \frac{1}{2}z^{2}(x^{2}+y^{2}) - V(x^{2}+y^{2}),
\end{eqnarray}
where $\dot{x}$ and $\dot{y}$ are generalized velocities and $z$ denotes the generalized coordinate. The canonical conjugate momenta 
corresponding to the generalized coordinates are listed below
\begin{eqnarray}
P_{x}=\dot{x}+zy,\qquad 
P_{y}=\dot{y}-zx,\qquad
P_{z}=0.
\end{eqnarray}
The Hamiltonian of the system can be obtained in the following fashion
\begin{eqnarray}
\tilde{H}=\frac{P_{x}}{2}^{2}+\frac{P_{y}}{2}^{2}-z(P_{x}y-P_{y}x)+V(x^{2}+y^{2}).
\end{eqnarray}
Before studying the system in the Faddeev-Jackiw formalism, we express the Lagrangian in first-order form where no quadratic 
time derivative terms are present, as  
\begin{eqnarray}
\tilde{L}^{(0)}_{f}=P_{x}\dot{x}+P_{y}\dot{y}-{\tilde V}^{(0)},
\end{eqnarray}
with
\begin{eqnarray}
\tilde{V}^{(0)}=\frac{P_{x}}{2}^{2}+\frac{P_{y}}{2}^{2}+z(P_{y}x-P_{x}y)+V(x^{2}+y^{2}).
\end{eqnarray}
The symplectic equations of motion derived in terms of symplectic matrix $\tilde{f}_{ij}^{(0)}$ are given as
\begin{eqnarray} \label{eqm_car}
\tilde{f}_{ij}^{(0)}\dot{\tilde{\zeta}}^{j} &=& \frac{\partial \tilde{V}^{(0)}(\tilde{\zeta})}{\partial \tilde{\zeta}^{i}}.
\end{eqnarray}
Here, we identify the set of symplectic variables in the theory as;
\begin{eqnarray}
\tilde{\zeta}^{(0)} = \left\{x, P_{x}, y, P_{y}, z \right\},
\end{eqnarray}
with the components of symplectic one-form, calculated according to \eqref{f_ij}, are 
\begin{eqnarray}
\tilde{a}^{(0)}_{x}=P_{x}, \quad \tilde{a}^{(0)}_{P_{x}}=0, \quad \tilde{a}^{(0)}_{y}=P_{y}, \quad \tilde{a}^{(0)}_{P_{y}}=0, 
\quad \tilde{a}^{(0)}_{z}=0.
\end{eqnarray}
Now, we determine the symplectic two-form matrix as (cf. \eqref{f_ij}) 
\begin{eqnarray}
\tilde{f}_{ij}^{(0)} =
\begin{pmatrix}
0 & -1 &0 &0 &0 \\
1 &0 &0 &0 &0 \\
0 &0 &0 &-1 &0\\
0 &0 &1 &0 &0\\
0 &0 &0 &0 &0
\end{pmatrix},
\end{eqnarray}
which is obviously a singular matrix. So it indicates the presence of constraints in the model. The zero-mode of the above symplectic matrix 
is obtained as $(\tilde \nu^{(0)})^{T}=(0,0,0,0, \tilde v^{z})$, where $\tilde v^{z}$ is an arbitrary constant. As Faddeev-Jackiw formalism 
suggests that the zero modes of 
$\tilde f_{ij}^{(0)}$ give rise to the constraints in the system, so from the equations of motion, zero mode of symplectic matrix produces 
\begin{eqnarray}
\tilde{\Omega}^{(0)}=(\tilde \nu^{(0)})^{T}\frac{\partial \tilde{V}^{(0)}(\tilde{\zeta})}{\partial \tilde{\zeta}^{(0)}} = 0, \label{41}
\end{eqnarray}
where
\begin{eqnarray}
\frac{\partial \tilde{V}^{(0)}(\tilde{\zeta})}{\partial \tilde{\zeta}^{(0)}} \; = \; 
\begin{pmatrix}
zP_{y}+2Vx\\
P_{x}-zy\\
-zP_{x}+2Vy\\
P_{y}+zx\\
xP_{y}-yP_{x}
\end{pmatrix}. \label{42}
\end{eqnarray}
Now, with the aid of (\ref{41}) and (\ref{42}), we obtain the following constraint in the system
\begin{eqnarray}
\tilde{\Omega}^{(0)}= \tilde \nu^{z}(xP_{y}-yP_{x}) =0.
\end{eqnarray} 
Furthermore, we make use of modified Faddeev-Jackiw formalism in order to deduce new constraints in the theory. The consistency condition of 
 obtained constraint ($\tilde{\Omega}^{(0)}$) gives rise to the following condition
\begin{eqnarray}
\dot{\tilde{\Omega}}^{(0)} = \frac{\partial{\tilde{\Omega}^{(0)}}}{\partial \tilde{\zeta}^{i}}\dot{\tilde{\zeta}}^{i} =0. \label{car_cons}
\end{eqnarray}
Combining  \eqref{eqm_car} and \eqref{car_cons}, we have
\begin{eqnarray}
\tilde{f}_{kj}^{(1)} \dot{\tilde{\zeta^{j}}} = \tilde{Z}_{k}(\tilde{\zeta}), \label{car_f_kj}
\end{eqnarray}
where
\begin{eqnarray}
\tilde{f}_{kj}^{(1)}=
\begin{pmatrix}
\tilde{f}_{ij}^{(0)}\\[6pt]
\frac{\partial{\tilde{\Omega}^{(0)}}}{\partial \tilde{\zeta}^{i}}
\end{pmatrix}
, \qquad \tilde{Z}_{k}(\tilde{\zeta})=
\begin{pmatrix}
\frac{\partial \tilde{V}^{(0)}(\tilde{\zeta})}{\partial \tilde{\zeta}^{i}}\\[6pt]
0
\end{pmatrix}.
\end{eqnarray}
Thus, we obtain the following symplectic matrix
\begin{eqnarray}
 \tilde{f}_{kj}^{(1)} =
\begin{pmatrix}
0 &-1 &0 &0 &0 \\
1 &0 &0 &0 &0 \\
0 &0 &0 &-1 &0\\
0 &0 &1 &0 &0\\
0 &0 &0 &0 &0\\
P_{y} &-y &-P_{x} &x &0
\end{pmatrix}.
\end{eqnarray}
This symplectic matrix is a non-square matrix. The calculation of zero-mode of the above matrix gives 
$(\tilde \nu^{(1)})^{T}=(-\alpha y,-\alpha P_{y},\alpha x,\alpha P_{x},\tilde v_{1}^{z},\alpha)$, where $\alpha$ and $\tilde v_{1}^{z}$ are arbitrary 
constants. 
Multiplying $(\tilde \nu^{(1)})^{T}$ to \eqref{car_f_kj} gives rise to new constraints in the theory.
\begin{eqnarray}
(\tilde \nu^{(1)})^{T} \tilde{Z}_{k}|_{\tilde{\Omega}^{(0)}=0} = 0 \; \Longrightarrow \; \tilde v_{1}^{z}(xP_{y}-yP_{x})|_{\tilde{\Omega}^{(0)}=0} = 0.
\end{eqnarray}
Thus, we obtained identity. So there are no further constraints present in the theory. Now, following Faddeev-Jackiw formalism, 
we introduce this constraint into the Lagrangian with the help of Lagrange multiplier $(\lambda)$, as
\begin{eqnarray}
\tilde{L}_f^{(1)}=P_{x}\dot{x}+P_{y}\dot{y}+(xP_{y}-yP_{x})\dot{\lambda}-\tilde{V}^{(1)},
\end{eqnarray}
where
\begin{eqnarray}
\tilde{V}^{(1)}=\tilde{V}^{(0)}|_{(xP_y - yP_x) = 0} \; = \; \frac{P_{x}}{2}^{2}+\frac{P_{y}}{2}^{2}+V(x^{2}+y^{2}).
\end{eqnarray}
The first-iterated set of symplectic variables are chosen to be
\begin{eqnarray}
\tilde{\zeta}^{(1)}= \left\{x,P_{x},y,P_{y},\lambda\right\},
\end{eqnarray}
and accordingly symplectic one-forms are being calculated as listed below 
\begin{eqnarray}
\tilde{a}^{(1)}_{x}=P_{x}, \quad \tilde{a}^{(1)}_{P_{x}}=0, \quad \tilde{a}^{(1)}_{y}=P_{y}, \quad \tilde{a}^{(1)}_{P_{y}}=0, 
\quad \tilde{a}^{(1)}_{\lambda}=xP_{y}-yP_{x}.
\end{eqnarray}
Thus, we acquire the first-iterated symplectic two-form matrix 
\begin{eqnarray}\label{54}
\tilde{f}_{ij}^{(1)} =
\begin{pmatrix}
0 &-1 &0 &0 &P_{y} \\
1 &0 &0 &0 &-y \\
0 &0 &0 &-1 &-P_{x}\\
0 &0 &1 &0 &x\\
-P_{y} &y &P_{x} &-x &0
\end{pmatrix}.
\end{eqnarray}
The matrix $ \tilde{f}_{ij}^{(1)}$ is a singular matrix. So the presence of singular symplectic matrix and absence of new constraints in the theory 
suggest that the system has a gauge symmetry.

As far as the quantization of Christ-Lee model, in Cartesian coordinates, is concerned we choose a gauge $y = 0$  \cite{13, 15}.
We incorporate this gauge condition into the first-iterated Lagrangian by means of a Lagrange multiplier $(\rho)$, as 
\begin{eqnarray} \label{car_L2}
\tilde{L}_f^{(2)}=P_{x}\dot{x}+P_{y}\dot{y}+(xP_{y}-yP_{x})\dot{\lambda}+y\dot{\rho}-\tilde{V}^{(2)},
\end{eqnarray}
where
\begin{eqnarray}
\tilde{V}^{(2)}=\tilde{V}^{(1)}|_{y=0} \; = \; \frac{P_{x}}{2}^{2}+\frac{P_{y}}{2}^{2}+V(x^{2}).
\end{eqnarray}
Now the second-iterated set of symplectic variables are
\begin{eqnarray}
\tilde{\zeta}^{(2)}= \left\{x, P_{x}, y, P_{y}, \lambda, \rho \right\}.
\end{eqnarray}
The elements of corresponding symplectic one-forms are given by
\begin{eqnarray}
\tilde{a}^{(2)}_{x}=P_{x}, \quad \tilde{a}^{(2)}_{P_{x}}=0, \quad \tilde{a}^{(2)}_{y}=P_{y}, \quad \tilde{a}^{(2)}_{P_{y}} = 0, 
\quad \tilde{a}^{(2)}_{\lambda}=xP_{y}-yP_{x},\quad \tilde{a}^{(2)}_{\rho} = y.
\end{eqnarray}
Thus, we obtain the second-iterated symplectic matrix as
\begin{eqnarray}
 \tilde{f}_{ij}^{(2)} =
\begin{pmatrix}
0 &-1 &0 &0 &P_{y} &0\\
1 &0 &0 &0 &-y &0\\
0 &0 &0 &-1 &-P_{x} &1\\
0 &0 &1 &0 &x &0\\
-P_{y} &y &P_{x} &-x &0 &0\\
0 &0 &-1 &0 &0 &0
\end{pmatrix}.
\end{eqnarray}
Here, the second-iterated symplectic matrix $ \tilde{f}_{ij}^{(2)}$ turns out to be non-singular and thus we can find its 
inverse in a straightforward manner as
\begin{eqnarray}
(\tilde{f}_{ij}^{(2)})^{-1}=
\begin{pmatrix}
0 &1 &0 &\frac{y}{x} &0 &\frac{y}{x} \\[6pt]
-1 &0 &0 &\frac{P_{y}}{x} &0 &\frac{P_{y}}{x}\\[6pt]
0 &0 &0 &0 &0 &-1\\[6pt]
-\frac{y}{x} &-\frac{P_{y}}{x}&0 &0 &-\frac{1}{x} &-\frac{P_{x}}{x}\\[6pt]
0 &0 &0 &\frac{1}{x} &0 &\frac{1}{x}\\[6pt]
-\frac{y}{x} &-\frac{P_{y}}{x} &1 &\frac{P_{x}}{x} &-\frac{1}{x} &0
\end{pmatrix}.
\end{eqnarray}
Moreover, the elements of $(\tilde{f}_{ij}^{(2)})^{-1}$ give the basic brackets existing in the theory. We have listed them as below
\begin{eqnarray} \label{bkts}
&&\{x,P_{x}\}=1=-\{P_{x},x\}, \quad \{x,P_{y}\}=\frac{y}{x}=-\{P_{y},x\},\\ \nonumber
&&\{P_{x},P_{y}\}=\frac{P_{y}}{x}=-\{P_{y},P_{x}\},
\quad \{P_{y},\lambda\}=-\frac{1}{x}=-\{\lambda,P_{y}\},\\ \nonumber
&&\{x,\rho\}=\frac{y}{x}=-\{\rho,x\},
\quad \{y,\rho\}=-1=-\{\rho,y\},\\ \nonumber
&&\{P_{x},\rho\}=\frac{P_{y}}{x}=-\{\rho,P_{x}\},
\quad \{P_{y},\rho\}=-\frac{P_{x}}{x}=-\{\rho,P_{y}\},\\ \nonumber
&&\{\lambda,\rho\}=\frac{1}{x}=-\{\rho,\lambda\}. 
\end{eqnarray}

As we have already pointed out the existence of gauge symmetry in the theory. We can obtain these gauge symmetries from the 
zero mode of the respective singular symplectic matrix (cf. (\ref{54})) which acts as the generators of the gauge transformations. 
The zero mode of the symplectic matrix $\tilde f_{ij}^{(1)}$ is given by
\begin{eqnarray}
(\tilde{\upsilon}^{(1)})^{T} = (y, P_{y}, -x, -P_{x}, 1).
\end{eqnarray}
Thus, the gauge transformations $(\tilde{\delta})$ can be deduced as 
\begin{eqnarray}
\tilde{\delta}\tilde{\zeta}_{k}^{(1)} = \tilde{\upsilon}_{k}^{(1)}\kappa,
\end{eqnarray}
where $\tilde{\zeta}_{k}^{(1)}$ is the set of all symplectic variables and $\kappa(t)$ represents the time-dependent infinitesimal 
parameter of gauge transformation. Therefore, in Cartesian coordinates, the gauge transformations can be explicitly given as 
\begin{eqnarray}
\tilde \delta x = y\kappa(t), \quad \tilde\delta P_{x} = P_{y}\kappa(t), \quad \tilde \delta y = -x\kappa(t), 
\quad \tilde \delta P_{y} = -P_{x}\kappa(t), \quad \tilde \delta z = -\dot{\kappa}(t).
\end{eqnarray}
It is straightforward to check that our first-order Lagrangian 
remains invariant under this transformation (i.e. $\tilde \delta{L_{f}^{(0)}}=0$).

\subsection{Interpretation of Lagrange Multipliers}
Here, in Faddeev-Jackiw formalism for Christ-Lee model in Cartesian coordinates we obtained basic brackets amongst  
Lagrange multipliers and physical coordinates (cf. (\ref{bkts}) above). As we have already mentioned 
that the brackets involving Lagrange multipliers do not exist in Dirac formalism. 
Similarly, as discussed in the previous section, we obtain a new insight about these Lagrange multipliers from
the symplectic equations of motion as written from \eqref{car_L2}
\begin{eqnarray}
\tilde{f}_{ij}^{(2)}\dot{\tilde{\zeta}}^{(2)j} = \frac{\partial \tilde{V}^{(2)}(\tilde{\zeta})}{\partial \tilde{\zeta}^{(2)i}},
\end{eqnarray}
\begin{eqnarray} \label{car_f_ij2}
\begin{pmatrix}
0 &-1 &0 &0 &P_{y} &0\\
1 &0 &0 &0 &-y &0\\
0 &0 &0 &-1 &-P_{x} &1\\
0 &0 &1 &0 &x &0\\
-P_{y} &y &P_{x} &-x &0 &0\\
0 &0 &-1 &0 &0 &0
\end{pmatrix} 
\begin{pmatrix}
\dot{x}\\
\dot{P_{x}}\\
\dot{y}\\
\dot{P_{y}}\\
\dot{\lambda}\\
\dot{\rho}
\end{pmatrix} =
\begin{pmatrix}
2Vx\\
P_{x}\\
0\\
P_{y}\\
0\\
0
\end{pmatrix}.
\end{eqnarray}
From above relation \eqref{car_f_ij2}, we obtain
\begin{eqnarray}
&& \dot{\lambda} \; = \; \frac{2Vx + \dot P_x}{P_y} \; = \; \frac{\dot x - P_x}{y} \; = \; \frac{P_y - \dot y}{x}, \nonumber\\
&& \dot \rho \; = \; \dot P_y + \frac{2VxP_x}{P_y} + \frac{P_x \dot P_x} {P_y} \; = \; \dot P_y + \frac{\dot x P_x}{y} - \frac{P^2_x}{y} 
\; = \; \dot P_y + \frac{P_x P_y}{x} - \frac{\dot y P_x}{x},
\end{eqnarray}
here the Lagrange multipliers can be expressed as the combination of generalized coordinates and generalized momenta.

\section{Conclusions}
In our present investigation, we have derived the constraints present in the Christ-Lee model and performed quantization {\it \`{a} la} 
Faddeev-Jackiw in both the polar and Cartesian coordinates. However, to obtain new constraints, we have used modified Faddeev-Jackiw method 
which makes use of the consistency condition of constraints analogous to Dirac-Bergmann approach.  
Even after incorporating all the obtained constraints into the Lagrangian through Lagrange multiplier the resulting two-form 
symplectic matrix turned out to be singular. This has clearly indicated that the underlying theory is a gauge theory. Thus, to quantize
the theory, we have appropriately chosen the gauge conditions which, in turn, made the two-form symplectic matrix to be non-singular. 
Thus, we obtained basic brackets from the inverse of this non-singular symplectic matrix in both the coordinate systems. 
Further, we inferred that the zero mode of the symplectic matrix gives the form of gauge transformations present in the theory. Finally,
we provided new interpretation of Lagrange multipliers in terms of physical coordinates of the system.

It will be interesting to figure out constraint structure and quantization of fractional Christ-Lee model within the framework of modified 
Faddeev-Jackiw formalism. Another captivating venture is to study this model within the framework of BRST and superfield formalism. 
These issues are presently being investigated and our results will be reported elsewhere \cite{20}.
\\

\noindent 
{\Large \bf Acknowledgments:} The support from FRG scheme of National Institute of Technology Calicut is thankfully acknowledged. 
Enlightening comments by esteemed Reviewers are gratefully acknowledged, too.

\end{document}